\documentclass[12pt, draftclsnofoot,onecolumn]{IEEEtran}
\usepackage{cite}
\usepackage{graphicx}
\usepackage{amsmath}
\usepackage{booktabs}
\usepackage{amsfonts}
\usepackage{balance}
\usepackage{fancyhdr}
\usepackage{graphicx,times,cite,amssymb,epsfig,amsthm,color}
\usepackage{algorithm}
\usepackage{algpseudocode}
\usepackage{cite}
\usepackage{graphicx}
\usepackage{amsmath}
\usepackage{amsfonts}
\usepackage{amssymb}
\usepackage{subcaption}
\usepackage{xcolor}

\newtheorem{theorem}{Theorem}

\newtheorem{lemma}{Lemma}

\newtheorem{corollary}{Corollary}

\newtheorem{proposition}{Proposition}

\newtheorem{conjecture}{Conjecture}

\newtheorem{sketch}{Sketch of Proof}

\begin{document}

\title{Frequency Permutations \\for Joint Radar and Communications}

\author{\IEEEauthorblockN{Rajitha Senanayake,~\IEEEmembership{Member,~IEEE,}
 Peter Smith,~\IEEEmembership{Fellow,~IEEE,}\\
  Tian Han,~\IEEEmembership{Student Member,~IEEE,}
   Jamie Evans,~\IEEEmembership{Senior Member,~IEEE,}\\
     William Moran,~\IEEEmembership{Member,~IEEE,} and Robin Evans,~\IEEEmembership{Life Fellow,~IEEE.}}
}

\maketitle

\begin{abstract}
This paper presents a new joint radar and communication technique based on the classical stepped frequency radar waveform.
The randomization in the waveform, which is achieved by using permutations of the sequence of frequency tones, is
utilized for data transmission. A new signaling scheme is proposed in which the mapping between incoming data and waveforms is performed based on an efficient combinatorial transform called the Lehmer code. Considering the optimum maximum likelihood (ML) detection, the union bound and the nearest neighbour approximation on the communication block error probability is derived for communication in an additive white Gaussian noise (AWGN) channel. The results are further extended to incorporate the Rician fading channel model, of which the Rayleigh fading channel model is presented as a special case.  Furthermore, an efficient communication receiver implementation is discussed based on the Hungarian algorithm which achieves optimum performance with much less operational complexity when compared to an exhaustive search.
From the radar perspective, two key analytical tools, namely, the ambiguity function (AF) and the Fisher information matrix are derived.
Furthermore, accurate approximations to the Cramer-Rao lower bounds (CRLBs) on the delay and Doppler estimation errors are derived based on which the range and velocity estimation accuracy of the waveform is analysed.
Numerical examples are used to highlight the accuracy of the analysis and to illustrate the performance of the proposed waveform.

\end{abstract}


\IEEEpeerreviewmaketitle
\section{Introduction}
Traditionally, wireless communications and radar sensing have been designed and developed in isolation addressing domain specific challenges.
Recently, there has been significant research interest in the cooperation and co-design of the two systems \cite{Sturm11,Hassanien16,Chiriyath17,Zheng19,Zhang21,Han13,Liu20}. This is primarily driven by the similarities of the two systems in terms of  signal processing algorithms, devices and  system architecture \cite{Zhang21}. With wireless communication systems starting to use millimeter-wave (mmWave) frequencies, which are traditionally used in radar systems, such joint radar and communication systems could also help increase the spectrum utilization. Its potential applies to a range of emerging applications including, autonomous vehicles \cite{Kumari18}, unmanned aerial vehicles \cite{Chen20}, the Internet of Things (IoT) \cite{Cianca17} and defence applications where communications and sensing both play important roles \cite{Ma20}.



\subsection{Related work}

The main challenge in developing such a joint system lies in finding a new waveform that is capable of simultaneously transmitting information and performing radar sensing. Existing literature proposes several new techniques to achieve this convergence. In \cite{Zheng19,Zhang21,Han13}, excellent reviews of joint radar and communication techniques are provided.

Several papers have proposed communication-centric designs in which traditional communication waveforms are exploited for radar sensing purposes.
In \cite{Kumari18}, a new waveform suitable for long-range radar (LRR) is proposed based on  the IEEE 802.11ad-based waveform which is traditionally used for  wireless local area networks (WLANs).
The special structure of repeated Golay sequences in the preamble of the 802.11ad-based waveform, which has good correlation properties, is exploited to estimate the radar parameters. In \cite{Daniels18}, the IEEE 802.11p waveform, which is used for  dedicated short-range communication (DSRC) in vehicles, is considered for radar range estimation of the closest target. The paper also presents an alternative to forward collision detection in vehicles based on the communication waveform. In \cite{Zeng19}, new algorithms based on wireless communication standards such as IEEE 802.11ad and IEEE 802.11p have been proposed for radar range and velocity estimation.

Orthogonal frequency division multiplexing (OFDM) waveforms, which incorporate frequency diversity, have also been proposed in \cite{Garmatyuk06,Guo14,Turlapaty14}, where the target range and Doppler is extracted from the reflected OFDM signal. The OFDM waveform also has similarities to phase coded radar. However, due to the addition of subcarrier components it introduces a high peak-to-average power ratio (PAPR) which is problematic in typical radar operation as it requires power amplifiers with a large linear range. While some attempt has been made to improve this by randomizing the carrier phase and amplitude \cite{Mozeson03}, the transmit power amplifier is not used efficiently from a radar perspective.
Some papers have proposed the use of cyclic prefixed orthogonal frequency division multiplexing (CP-OFDM) which has limitations involving the length of the cyclic prefix \cite{Braun10}.
Similar to OFDM, spread spectrum waveforms such as  direct sequence spread spectrum (DSSS) have also been proposed for joint radar and communications systems \cite{Shaojian06}.

From the opposite perspective, several papers have focused on using conventional radar waveforms for joint radar and communication purposes. In \cite{Bocquet10,Richmond16}, pulse position modulation and pulse amplitude modulation are considered to embed data in pulsed radar signals. The popular linear frequency modulated (LFM) waveform is considered in \cite{Roberton03}, where separate up and down chirps are used for radar and communication pulses, respectively. In this work, communication data rates are further enhanced by using $\pi/4$-Differential Quadrature Phase Shift Keying (DQPSK) modulation. The LFM waveform is also considered in \cite{Blunt10}, resulting in a low communication symbol rate that corresponds only to the chirp rate. Recently, a chirp-based $M$-ary Frequency Shift Keying (chirp-$M$-FSK) modulation method has been proposed in \cite{Dwivedi19}, together with the frequency modulated continuous wave (FMCW) radar waveform. 
Embedding data into  traditional radar waveforms is challenging due to the lack of randomness in the available parameters. Also, when adjusting waveform parameters it is important to make sure radar performance is not compromised.
None of the above works  characterise the effect of embedding data on the radar performance in-terms of key analytical tools in radar waveform design such as the \textit{ambiguity function} and the Fisher information matrix. As such, analysis of the behavior of  new waveforms under important phenomena such as radar clutter is lacking.

\subsection{Motivation and Contributions}
In this paper, we focus on the stepped frequency radar waveform, which, to the best of our knowledge, has not been considered for this joint operation before, and propose a new method to randomize the frequency tones to modulate data. In order to make the paper complete and clearly readable to both the communications and radar communities we have conducted a rigorous analytical investigation on a number of important performance measures.  While some of these measures are considered in the literature they are not always rigorously developed. We hope that the level of detail presented here is helpful and valuable to the reader.

Radar waveforms consisting of a sequence of step changes in frequency are popular in many radar applications and are being used extensively in the emerging automotive radar application \cite{Skaria19,Hourani18,Hourani17}. In a stepped frequency waveform, the available bandwidth is divided into equally spaced frequency tones separated by a constant step $\Delta f$ Hz. The most common stepped frequency waveform employs a linear stepping pattern where the frequency of each pulse is increased by $\Delta f$ from the preceding pulse \cite{richards14}. Costas codes, on the other hand, arrange the frequency tones according to a special sequence that results in an extremely good radar waveform with zero side-lobes \cite{Levanon04}. However, the number of available Costas codes are limited and it does not provide enough randomization to be used in a joint radar communication system. Interestingly, in \cite{Skaria19,Hourani18}, a pseudo random stepped frequency radar is considered where the transmitted waveforms consist of a train of short tones with frequencies defined according to a pseudo-random sequence. The randomization in the sequence of tones appears to be more suitable for the automotive radar application as it can significantly reduce the interference arising from a large number of radars operating in close proximity \cite{Skaria19}.

In this paper, we exploit this randomization in frequency to send data. We focus on a random stepped frequency radar waveform based on \textit{Lehmer codes}, in which the transmitted waveform consists of a train of short tones with frequencies defined according to a permutation of the sequence of tones. As such, a set of $M$ frequency tones generates $M!$ different waveforms - all suitable for typical radar applications. These waveforms also maintain a constant amplitude resulting in a PAPR of unity which is ideally suited to radar operation. Our novel contributions are detailed as follows:
\begin{itemize}
  \item We propose a novel joint radar and communication technique in which the incoming data is modulated based on the selection of the frequency permutation in the transmitted waveform. We provide an efficient implementation for the mapping between the incoming data and the corresponding waveform based on a combinatorial transform called the Lehmer code.
  \item  From a communication perspective, we consider optimum maximum likelihood (ML) detection and analyse the baseband communication in additive white Gaussian noise (AWGN) channels to provide a union bound and a nearest neighbour approximation on the  block error probability of the proposed signaling scheme. Then, we extend the analysis to consider Rician fading and provide closed-form analytical expression for the same performance measures. The Rayleigh channel model is also considered as a special case.
  \item We provide an efficient communication receiver implementation based on the Hungarian algorithm which always results in the optimum solution. Compared to the exhaustive search which has a time complexity of $O(M!)$, the Hungarian algorithm has a worst case complexity of $O(M^3)$, where $M$ is the number of frequency tones in the waveform.
  \item  From a radar perspective, we derive the ambiguity function and the Fisher information matrix of the proposed waveform. Based on the ambiguity function, we analyse the behavior of the waveform in relation to the matched filter output and provide a detailed discussion on the overall structure of the ambiguity function. Based on the Fisher information matrix we derive approximate Cramer-Rao Lower Bounds (CRLBs) on the delay and Doppler estimation error and  illustrate the effect of different parameters on radar range and Doppler resolution.
\end{itemize}
Numerical examples are provided to support the discussion and also to illustrate the accuracy of the analysis.
The remainder of the paper is organized as follows. In Section \ref{sys}, the system model and the waveform of the novel joint radar and communications system are described. The communication and radar performance of the Lehmer code based random stepped frequency radar waveform is analysed in Section \ref{comm} and \ref{radar}, respectively. Simulation  results  are  provided  in  Section  \ref{NR}  with conclusions and future research directions  provided in Section \ref{con}.





\section{Joint Radar and Communications System }\label{sys}
\subsection{System Model}
\begin{figure}[t]
    \centerline{\includegraphics[width=8cm,height=4cm]{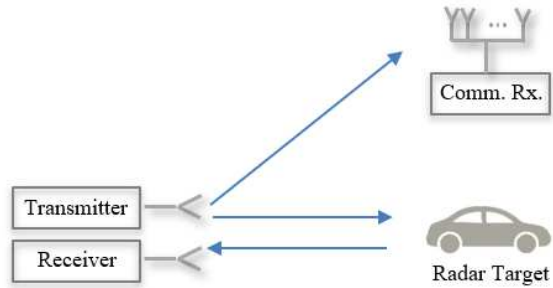}}
    \caption{Illustration on the joint radar communication system model.
    }\label{fig0}
\end{figure}

Consider the  system model illustrated in Fig. \ref{fig0}, where the transmitter sends a joint waveform for both communication and radar systems.
The return signal from the radar target is received and processed by the radar receiver to estimate the range and the velocity of the target. The signal received at the communications receiver, which is equipped with multiple antennas, is processed to detect the transmitted information.

\subsection{Joint Radar and Communications Waveform }

\begin{figure}[t]
    \centerline{\includegraphics[width=8.5cm,height=3.4cm]{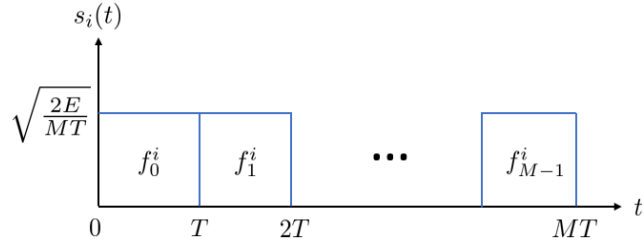}}
    \caption{The envelope of the $i$-th Lehmer code based random stepped frequency waveform.
    }\label{fig1}
\end{figure}

Consider $M$ equally spaced frequency tones $f_0, f_1, \ldots, f_{M-1}$. 
Based on the fact that there are $M!$ permutations of this sequence of $M$ frequencies we formulate $M!$ stepped frequency radar waveforms. In each waveform a given frequency is used only once. As illustrated in Fig. \ref{fig1}, the $i$-th waveform is generated using $M$ pulses, each $T$ seconds long and, the frequencies follow the $i$-th permutation of the set of $M!$ permutations with $f^i_m$ denoting the $m$-th frequency of the $i$-th permutation. As such, the complex baseband representation of the $i$-th waveform is given by
\begin{align}\label{eq1}
s_i(t) = \sqrt{\frac{E}{MT}}\sum_{m = 0}^{M-1} s_p(t - mT)\exp(2\pi f^i_m(t - mT)),
\end{align}
where
\begin{align}\notag
s_p(t) = \left\{
\begin{array}{ll}1~~~~~~~~0 \leq t \leq T\\
0~~~~~~~~\text{otherwise}.
\end{array}\right.
\end{align}
$T$ denotes the pulse duration and $E$ is the signal energy satisfying
\begin{align}\notag
\int^{MT}_0s_i^2(t)dt = E,~~~~~~~~i = 0, 1, \ldots, M-1.
\end{align}
We also assume that the frequencies are orthogonal. Thus, the separation between two frequency tones $\Delta f = n/T$, where $n$ is an integer.
Since all frequencies are permutations of the $M$ frequency tones, the set of $M!$ sequences are given by
\begin{align}\label{eq14}
&\psi_1 = \{f_0, \ldots, f_{M-2}, f_{M-1}\},\notag\\
&\psi_2 = \{f_0, \ldots, f_{M-1}, f_{M-2}\},\notag\\
&~~~~~\vdots
\notag\\
&\psi_{M!} = \{f_{M-1}, \ldots, f_1, f_0\}\notag.
\end{align}
We note that all $M!$ waveforms generated by the above permutations  maintain a constant amplitude. This, in contrast to OFDM, results in a PAPR of unity which is suitable for typical radar operation.  
Moreover, these waveforms maintain a short individual pulse width $T$ which results in good delay resolution. A good Doppler resolution is obtained by changing the frequency from one pulse to another, thus, utilizing  a total bandwidth of $M \Delta f$ when considering the waveform as a whole. Due to the randomness in frequency,  all $M!$ waveforms generated by the permutations are suitable for the automotive radar application \cite{Hourani17}. We embed data in the selection of the waveform by maintaining a direct mapping between the data symbol and the selected waveform. In other words, we take $\log_2{M!}$ number of bits and depending on the symbol it represents, select the waveform to be transmitted.

\subsection{Use of Lehmer Codes}
Since we have $M!$ possible waveforms, one could argue that for large $M$, the implementation of  the mapping between incoming data and the corresponding waveform may not be practical as it requires a large lookup table. However, it is important to identify that the the use of a lookup table is unnecessary for the implementation of the proposed signaling scheme. The mapping between incoming data and the corresponding waveform can be efficiently implemented using a combinatorial transform  called the \textit{Lehmer code} \cite{Lehmer60}.
With Lehmer codes, we first present the natural number of the incoming data symbol in the factorial number system, which is also known as the factoradic form. Then, we exploit the natural mapping between the integers $0, 1, \ldots, M!-1$ and permutations of $M$ elements in lexicographic order 
to map incoming data to a unique permutation of frequency tones \cite{Knuth73}.
At the communications receiver, a similar decoding procedure is followed to map the detected permutation of frequency tones to the corresponding data symbol. This procedure is a straightforward mapping between integers and permutations and can be efficiently implemented as it only involves simple operations such as selection from a sequence and deletion from it. For more details on mapping and demapping using Lehmer codes please refer to \cite{Knuth73}.

\section{Communication Performance Analysis}\label{comm}
In this section, we focus on the communications performance of the proposed joint radar and communications system. We consider
ML detection at the communications receiver and analyse the error probability performance. 

\subsection{Maximum Likelihood Detection}
Consider a communication receiver with $N$ receive antennas.
The $N \times 1$  complex baseband received signal vector is given by
\begin{align}
\mathbf{r}(t) = \mathbf{h}s_i(t) + \mathbf{n}(t),
\end{align}
where $\mathbf{h}$ is the small-scale fading channel vector, $s_i(t) \in \{s_1(t), s_2(t), \ldots, s_{M!}(t)\}$  is one of the $M!$ possible waveforms bearing transmit data and $\mathbf{n}(t)$ is an AWGN vector where each element is complex Gaussian with zero mean and variance $N_0$.

Assuming that the channel vector $\mathbf{h}$ is known at the receiver, we consider a coherent ML detector and write
 write the detected symbol as
\begin{align}\label{eq3}
\hat{s}_i(t) = {\underset{{s_k}(t) \in \{{s}_1(t), {s}_2(t), \ldots, {s}_{M!}(t)\}} {\text{arg min}}} \int_0^{MT} |\mathbf{h}^H\mathbf{r}(t) - \mathbf{h}^H\mathbf{h}{s_k}(t)|^2 dt.
\end{align}
Using some straight forward mathematical manipulations, the ML detection rule in \eqref{eq3} can be reexpressed as
\begin{align}\label{eq4}
\hat{s}_i(t) = {\underset{{s}_k(t) \in \{{s}_1(t), {s}_2(t), \ldots, {s}_{M!}(t)\}}{\text{arg max}}} \textrm{Re}\left(\int_0^{MT}{s_k}^*(t)\mathbf{h}^H\mathbf{r}(t) dt\right),
\end{align}
where $\textrm{Re}(.)$ denotes the real part of the argument.

\subsection{Error Probability Performance}
Consider the hypothesis that $s_i(t)$, which corresponds to the the $i$-th sequence, is being transmitted. The sequence is correctly detected if $\hat{s}_i(t) = {s}_i(t)$. Thus, the probability of a correct decision is given by
\begin{align}\label{eq5}
P_c(i) = \textrm{Pr}\left[\xi_{ii} = {\underset{k \in \{1, 2, \ldots M!\}}{\textrm{max}}} \xi_{ik}\right],
\end{align}
where
\begin{align}\notag
\xi_{ik} = \textrm{Re}\left(\int_0^{MT}{s}^*_k(t)\mathbf{h}^H\mathbf{r}(t) dt\right)
\end{align}
denotes the decision variable.
The transmitted data is assumed to be equally likely. Thus, the a priori probabilities of the $M!$ sequences are $1/M!$, and the error probability performance of the sequence can be defined based on a block error rate as
\begin{align}\label{eq6}
P_e = \frac{1}{M!} \sum_{i = 1}^{M!} [1 - P_c(i)].
\end{align}
The exact expression of $P_e$ requires complex multi-dimensional integrals over the multivariate Gaussian density. These integrals cannot be computed in closed-form. Thus, we take a more tractable approach by considering the union bound which is given by
\begin{align}\label{eq7}
P_e \leq P_e^{UB} = \frac{1}{M!} \sum_{i = 1}^{M!} \sum_{k = 1, k \neq i}^{M!} P_{ik},
\end{align}
where $P_{ik}$ denotes the pairwise error probability of detecting the $k$-th sequence when the $i$-th sequence is transmitted. Please note that while the union bound is commonly used in literature, its derivation in the present context is not straight forward as we consider an unconventional method of modulating data based on the frequency permutation.
Let $!x$, given by
\begin{align}\label{eq8}
!x = x!\sum_{i = 0}^x\frac{(-1)^i}{i!},
\end{align}
denote the number of derangments of an $x$-element set \cite{Mehdi03}.
By analysing the $M!$ permutations of the sequence of $M$ frequency tones we note that
we have $!l \binom{M}{M - l}$ pairwise error probabilities where the frequency tones differ by $l$ positions. Using this observation and considering the symmetry in pairwise error probabilities, we can further simplify \eqref{eq7} as
\begin{align}\label{eq9}
P_e^{UB} = \sum_{l = 2}^{M}  !l\binom{M}{M - l} P_{1{k_l}},
\end{align}
where $k_l \in \{1, 2, \ldots, M\}$ is such that $\psi_1$ and $\psi_{k_l}$, which are the frequency sequences used to generate $s_1(t)$ and $s_{k_l}(t)$, respectively, differ by $l$ frequency tones.
Next we proceed to derive $P_{1k_l}$ which is the probability of detecting $s_{k_l}(t)$ when $s_1(t)$ is transmitted. 
 Using the ML detection rule we can write  $P_{1k_l}$ as
\begin{align}\label{eq10}
P_{1k_l} = \textrm{Pr}&\left[\textrm{Re}\left(\int_0^{MT}{s}^*_1(t)\mathbf{h}^H\mathbf{r}(t) dt\right) < \textrm{Re}\left(\int_0^{MT}{s}^*_{k_l}(t)\mathbf{h}^H\mathbf{r}(t) dt\right)\right].
\end{align}
Using some  mathematical manipulations \eqref{eq10} can be reexpressed as
 \begin{align}\label{eq102}
P_{1k_l} = \textrm{Pr}\left[ \frac{El}{M} \mathbf{h}^H\mathbf{h}  < N_2 - N_1\right],
\end{align}
where $N_1  = \textrm{Re}\left( \mathbf{h}^H\int_0^{MT}{s}^*_1(t)\mathbf{n}(t) dt \right)$ and  $N_2  = \textrm{Re}\left( \mathbf{h}^H\int_0^{MT}{s}^*_{k_l}(t)\mathbf{n}(t) dt \right)$. When conditioned on  channel fading, $N_2 - N_1$ has a Gaussian distribution with zero mean and variance $\frac{ElN_0}{M} \mathbf{h}^H\mathbf{h}$. Thus, we can write a simple expression for  $P_{1k_l}$  as
\begin{align}\label{eq11}
P_{1k_l} = \textrm{Pr}&\left[ \sqrt{\mathbf{h}^H\mathbf{h}} < \alpha_l Z \right],
\end{align}
where $\alpha_l = \sqrt{\frac{N_0M}{El}}$ and $Z \sim \mathcal{N}(0,1)$. 
Next, we proceed to solve \eqref{eq11} with respect to different wireless communication channel models.

\subsubsection{AWGN Channel}
To gain fundamental insights on the error performance, first, let us focus on   AWGN channels. Without loss of generality we assume a unit channel gain so that $\mathbf{h}^H\mathbf{h} = N$, which results in
\begin{align}\label{eq11d}
P_{1k_l} = \mathcal{Q}\left(\frac{\sqrt{N}}{\alpha_l}\right),
\end{align}
where $Q(.)$ is the Gaussian Q-function. Substituting \eqref{eq11d} into \eqref{eq9}  the union bound on the block error rate under AWGN channels can be written as
\begin{align}\label{eq19d}
P_e^{UB} = \sum_{l = 2}^{M}  !l\binom{M}{M - l}\mathcal{Q}\left(\frac{\sqrt{N}}{\alpha_l}\right).
\end{align}
While the union bound is commonly used in communications to provide a simple upper bound on the error probability performance, it has some limitations, especially for large $M$. The union bound considers all the pairwise error probabilities when calculating the error probability. Therefore,  as we increase $M$ the bound is not very accurate, especially in the low SNR regime. To avoid such limitations, in the following, we derive another approximation based on the nearest neighbour approximation \cite{Proakis08}. Under this approach, we only consider the pairwise error probabilities corresponding to the nearest neighbours. As the waveforms are generated using permutations of frequency tones we note that the nearest neighbours only differ by two frequency tones resulting in a new approximation given by
\begin{align}\label{eq20d}
P_e^{NN} =    \frac{M(M - 1)}{2} \mathcal{Q}\left(\frac{\sqrt{N}}{\alpha_2}\right),
\end{align}
where $M(M - 1)$ is the number of pairwise error probabilities where the frequency tones differ by two positions.

\subsubsection{Rician Fading Channel}
In the context of vehicular communication, fading in the propagation channel is inevitable due several reasons including the low position of antennas, large numbers of reflecting metallic surfaces surrounding the transmitters and the high mobility of transmitters and receivers \cite{bernado14}. To gain further insights into the effect of fading, next, we focus on the independent Rician fading model in which a line-of-sight (LOS) signal is present, along with other scatterers.
As such, we write the $N \times 1$ small scale fading channel vector as
\begin{align}\label{eq12}
\mathbf{h} = \sqrt{\frac{K}{K+1}}\Delta +  \sqrt{\frac{1}{K+1}}\mathbf{u},
\end{align}
where $\Delta$ denotes the complex LOS phase vector with the $i$-th element having the property $|\Delta_i|^2 = 1$, $\mathbf{u}$ denotes the scattered component vector with the $i$-th element $u_i \sim \mathcal{CN}(0,1)$. The strength of the LOS path is governed by the Rician factor given by $K$.
Based on \eqref{eq12} we can write  $\mathbf{h}^H\mathbf{h}$ as
\begin{align}\label{eq13}
\mathbf{h}^H\mathbf{h} = \frac{1}{2(K+1)}\chi_{2N}^2(2NK),
\end{align}
where $\chi_{2N}^2(2NK)$ denotes a non-central chi-squared distribution with $2N$ degrees of freedom and non-centrality parameter $2NK$. The main steps of the derivation of \eqref{eq13} is given in Appendix \ref{appenA}.
Substituting \eqref{eq13} into \eqref{eq11} we can write
\begin{align}\label{eq14c}
P_{1k_l} = \textrm{Pr}&\left[ \sqrt{\chi_{2N}^2(2NK)} < \sqrt{c_l} Z \right],
\end{align}
where $c_l = {2\alpha_l^2 (K+1)}$ and the probability in \eqref{eq14c} is taken both with respect to $\chi_{2N}^2(2NK)$ and $Z$.
Based on the CDF of the non-central chi-squared distribution \cite{Johnson70}, we can reexpress \eqref{eq14c} as
\begin{align}\label{eq15}
P_{1k_l} = \sum_{j = 0}^\infty\left(\frac{(NK)^je^{-NK}}{j!}\right)\textrm{Pr}\left[ \sqrt{\chi_{2N+2j}^2} < \sqrt{c_l} Z \right].
\end{align}
Noting that $Z$ has a standard normal Gaussian distribution and $\chi_{2N+2j}^2$ has a chi-squared distribution with $2N + 2j$ degrees of freedom we proceed to solve the above probability as
\begin{align}\label{eq16cc}
P_{1k_l} = \sum_{j = 0}^\infty\left(\frac{(NK)^je^{-NK}}{j!}\right)\int_{0}^\infty Q\left(\sqrt{\frac{2x}{c_l}}\right) \frac{x^{N + j - 1}e^{-x}}{\Gamma(N + j)} dx.
\end{align}
where $Q(.)$ is the Gaussian Q-function and $\Gamma(.)$ is the Gamma function \cite{grashteyn07}. Using  Craig's formula to reexpress the Gaussian Q-function we simplify the integral in \eqref{eq16cc} as
\begin{align}\label{eq17}
P_{1k_l} = \frac{1}{\pi}\sum_{j = 0}^\infty\left(\frac{(NK)^je^{-NK}}{j!}\right)\int_{0}^{\frac{\pi}{2}}\left(\frac{\sin^2{\theta}}{c_l^{-1} + \sin^2{\theta}}\right)^{N + j}d\theta,
\end{align}
which can be evaluated in closed-form using \cite[eq. 69]{Mallik02}  to produce
\begin{align}\label{eq18c}
P_{1k_l} = \sum_{j = 0}^\infty\left(\frac{(NK)^je^{-NK}}{j!}\right)\left(\frac{1}{2} + \sum_{n = 1}^{N+j}(-1)^n\binom{N+j}{n}\nu_n(c_l)\right),
\end{align}
where $\nu_n(c_l) = \frac{1}{2(1 + c_l)^{n - 1/2}}\sum_{q = 0}^{n - 1}\binom{n-1}{q}\binom{2q}{q}\left(\frac{c_l}{4}\right)^q$. Substituting \eqref{eq18c} into \eqref{eq9} the final closed-form expression for the union bound on the block error rate can be written as
\begin{align}\label{eq19c}
P_e^{UB} = \sum_{l = 2}^{M} \sum_{j = 0}^\infty !l\binom{M}{M - l}\left(\frac{(NK)^je^{-NK}}{j!}\right)\left(\frac{1}{2} + \sum_{n = 1}^{N+j}(-1)^n\binom{N+j}{n}\nu_n(c_l)\right).
\end{align}
Similar to the AWGN channel, the nearest neighbour approximation under the Rician fading channel model is given by
\begin{align}\label{eq20c}
P_e^{NN} =   \sum_{j = 0}^\infty \left(\frac{M(M - 1)(NK)^je^{-NK}}{2j!}\right)\left(\frac{1}{2} + \sum_{n = 1}^{N+j}(-1)^n\binom{N+j}{n}\nu_n(c_2)\right).
\end{align}
In Section \ref{NR}, we illustrate the accuracy of the above analysis using numerical examples and further discuss the error probability performance of the proposed signaling scheme under  Rician fading channels.

\subsubsection{Special Case - Rayleigh Fading Channel}
Next, we focus on the independent Rayleigh fading model which is a special case of  Rician fading. By setting $K = 0$ in \eqref{eq19c}, we can derive the union bound on the block error rate  as
\begin{align}\label{eq21}
P_e^{UB} = \sum_{l = 2}^{M}  !l\binom{M}{M - l}\left(\frac{1}{2} + \sum_{n = 1}^{N}(-1)^n\binom{N}{n}\nu_n(c_l)\right).
\end{align}
Similarly, the nearest neighbour approximation under Rayleigh distribution can be written as
\begin{align}\label{eq22}
P_e^{NN} =   \frac{M(M - 1)}{2}\left(\frac{1}{2} + \sum_{n = 1}^{N}(-1)^n\binom{N}{n}\nu_n(c_2)\right).
\end{align}
This provides an accurate approximation of the block error probability when there is no LOS signal present and the communication channel fading is Rayeigh distributed.
In Section \ref{NR}, we illustrate the accuracy of the above analysis using numerical examples.

\subsection{Communications Receiver Implementation}

We propose an efficient method for the implementation of the optimal communications receiver under the proposed modulation method.
An optimal receiver for the fading channel implements the ML decision rule, which is given in \eqref{eq4}, where the detected signal  maximizes correlation between the transmitted signal and the received signal correlated with the channel vector. This can be implemented using a correlation receiver (or a matched filter) using the relation
\begin{align}\label{eq122}
r_{nm} =  \textrm{Re}\left(\int_{(n-1)T}^{nT} \mathbf{h}^H\mathbf{r}(t)\phi_m(t - (n-1)T) dt\right),
\end{align}
where the basis function $\phi_m(t)$ is defined by
$\phi_m(t) =  \sqrt{\frac{2}{T}}s_p(t)\cos(2\pi f_m(t))$.
Based on \eqref{eq122}, we formulate the following matrix
\begin{align}\label{eq132}
\textbf{R} = (r_{nm}) \in \mathbb{R}^{M \times M},
\end{align}
in which the $nm$-th element, $r_{nm}$, represents the correlation between the received signal and the basis function $\phi_m(t - (n-1)T)$.
Now, the ML detection rule reduces to selecting the $M$ elements from $\textbf{R}$ in such a way that the selected $M$ elements are in $M$ different rows (because the frequencies are not repeated in a given waveform) and $M$ different columns (because only one frequency is used during a given pulse duration) and the sum of the $M$ elements is maximized.

We note that this optimization is an assignment problem which can be solved efficiently using the Hungarian algorithm \cite{kuhn55}. More specifically, we formulate the cost matrix by negating $\textbf{R}$ and follow the Hungarian algorithm to pick at most one element from each row and column in such a way that the summation is maximized. It is important to note that the Hungarian algorithm always results in the optimum solution. For the sake of completion, in Algorithm 1, we summarize the main steps of the Hungarian algorithm that we follow  in order to obtain the optimum solution. A working example of the algorithm is also provided in \ref{appenAA}. For further details please refer to \cite{kuhn55}.
Compared to the exhaustive search which has a time complexity of $O(M!)$, the Hungarian algorithm has a worst case complexity of $O(M^3)$.

\begin{algorithm}
\caption{Hungarian Algorithm for the Communications Receiver}
\begin{algorithmic}[1]
\State Negate the cost matrix $\textbf{R}$ to produce $(-\textbf{R})$
\State Subtract the row minimum from each row in $(-\textbf{R})$ and assign the resulting matrix to $\tilde{\textbf{R}}$
\State Subtract the column minimum from each column in $\tilde{\textbf{R}}$ and assign the resulting matrix to $\bar{\textbf{R}}$
\State $count = 0$
    \While{$count \not= M$}
        \State $count$ = minimum number of lines required to cover all zeros in $\bar{\textbf{R}}$.
        \\Note that covering a zero means putting a line through the row with that zero. The maximum number of lines that can be drawn through a row is one.
        \If{$count == M$}
        \State break;
        \Else
        \State $r_{min} =$ the smallest uncovered number in $\bar{\textbf{R}}$
        \State Update $\bar{\textbf{R}}$ by subtracting $r_{min}$ from all uncovered numbers  and add $r_{min}$ to all the elements that are covered twice
        \EndIf
    \EndWhile  \label{algo1}
    \State{The values of the optimal assignment are covered by the zeros in $\bar{\textbf{R}}$}
\end{algorithmic}
\end{algorithm}

\section{Radar Performance Analysis}\label{radar}
So far, we have considered the communications performance of the  Lehmer code based random stepped frequency radar waveform. In this section, we focus on the radar performance of this waveform and derive the two key analytical tools in radar waveform design, namely, the ambiguity function and the Fisher information matrix based on which we evaluate the side-lobe structure of the waveform and derive the CRLBs on velocity and range estimation. Using these analytical tools we fully characterize the delay and Doppler response of the radar waveform and discuss the effect of embedding data on the target range and velocity estimation.


\subsection{Ambiguity Function}
In radar waveform design, the time-frequency autocorrelation function of the transmitted waveform provides  a measure of the degree of similarity between the complex envelope of the transmitted waveform and a replica of it that is shifted in time and frequency \cite{vantrees01}.
The ambiguity function is defined as the magnitude of the time-frequency autocorrelation function and it fully characterizes the behavior of a radar waveform in relation to the matched filter output \cite{richards14,Rihaczek96}. More specifically, it describes the output of a matched filter when the signal input to the radar receiver is delayed by $\tau$ and Doppler shifted by $\omega$. 
The ambiguity function is useful for examining several radar parameters including, resolution, side-lobe structure and accuracy in both delay and Doppler.
In fact, the local accuracy of the delay and Doppler estimates are characterized by the behavior of the ambiguity function around the origin, while the global accuracy is characterized by its broader structure \cite{vantrees01}.



In order to derive the ambiguity function, we re-write the expression in \eqref{eq1}  as
\begin{align}\label{eq14}
s_i(t) = \sqrt{\frac{1}{MT}}\sum_{m = 0}^{M-1} s_p(t - mT)\exp [j\omega^i_m(t - mT)],
\end{align}
where $\omega^i_m = 2\pi  f^i_m$ and without loss of generality we have normalized the signal energy to one.
Based on \cite[eq. (4.30)]{richards14}, the time-frequency autocorrelation function, which is also known as the complex ambiguity function,  of the above waveform can be written using $s_i(t)$ and its complex conjugate $s_i^*(t)$ as
\begin{align}\label{eq15}
\hat{A}(\tau, \omega) = \int_{-\infty}^{\infty} s_i(s)s_i^*(s - \tau) \exp(j\omega s)ds,
\end{align}
where $\tau$ is the time delay relative to the expected matched filter peak output and $\omega$ is the Doppler mismatch.
Substituting \eqref{eq14} into \eqref{eq15} we can write
\begin{align}\label{eq16}
\hat{A}(\tau, \omega) = \frac{1}{MT} \sum_{m = 0}^{M - 1}&\sum_{n = 0}^{M - 1}\int_{-\infty}^{\infty} s_p(s - mT)s_p^*(s - \tau - nT)  e^{j (\omega s + \omega_m^i(s - mT) - \omega_n^i(s - \tau - nT))}ds.
\end{align}
Substituting $s' = s - mT$ and rearranging \eqref{eq16} we get
\begin{align}\label{eq16a}
\hat{A}(\tau, \omega) = \frac{1}{MT} &\sum_{m = 0}^{M - 1}\sum_{n = 0}^{M - 1}e^{j\omega mT}\int_{-\infty}^{\infty} s_p(s')s_p^*(s' - \tau - nT + mT) e^{j (\omega s' + \omega_m^is' - \omega_n^i(s' - \tau - nT+ mT))}ds'.
\end{align}
If the complex ambiguity function of the simple pulse $s_p(t)$ is represented by $\hat{A}_p(\tau, \omega)$, the integral in \eqref{eq16a} can be reexpressed as
\begin{align}\label{eq17}
\hat{A}(\tau, \omega) = \frac{1}{MT} \sum_{m = 0}^{M - 1}\sum_{n = 0}^{M - 1} & \hat{A}_p(\tau + nT - mT, \omega - \omega_n^i + \omega_m^i) e^{j (\omega mT - \omega_n^i((m - n)T - \tau))},
\end{align}
where $\hat{A}_p(\tau, \omega)$ is given by
\begin{align}\label{eq18}
\hat{A_p}(\tau, \omega) = \left\{
\begin{array}{ll} e^{j\omega \tau}\left(\frac{e^{j\omega T} - e^{-j\omega \tau}}{j\omega}\right)~~~~~~~~-T \leq \tau < 0\\
\frac{e^{j\omega T} - e^{j\omega \tau}}{j\omega}~~~~~~~~~~~~~~~~~~~~0 \leq \tau \leq T\\
0~~~~~~~~~~~~~~~~~~~~~~~~~~~~~~~\text{otherwise}.
\end{array}\right.
\end{align}
Then, the ambiguity function of the waveform is defined as the magnitude of $\hat{A}(\tau, \omega)$
\begin{align}\label{eq19}
{A}(\tau, \omega) = |\hat{A}(\tau, \omega)|.
\end{align}
The expression in \eqref{eq19} succinctly characterizes the behavior of the Lehmer code based stepped frequency radar waveform in the whole delay-Doppler plane. 

The matched filter output when there is no Doppler mismatch is given by the zero-Doppler response ${A}(\tau, 0)$, which can be found by setting $\omega = 0$ in \eqref{eq19} which results in
\begin{align}
{A}(\tau, 0) &= \left|\frac{1}{MT} \sum_{m = 0}^{M - 1}\sum_{n = 0}^{M - 1}  \hat{A}_p(\tau + (n - m)T,  - \omega_n^i + \omega_m^i)  e^{j (- \omega_n^i((m - n)T - \tau))}\right|.
\end{align}
Note that ${A}(\tau, 0)$ is the autocorrelation function of $s_i(t)$.
Similarly, the matched filter output when there is no delay is given by the zero-delay response ${A}(0, \omega)$, which can be found by setting $\tau = 0$ in  \eqref{eq19} which results in
\begin{align}
{A}(0, \omega) &= \left|\frac{1}{MT} \sum_{m = 0}^{M - 1}\sum_{n = 0}^{M - 1}  \hat{A}_p((n - m)T, \omega - \omega_n^i + \omega_m^i) e^{j (\omega mT - \omega_n^i((m - n)T))}\right|.
\end{align}
Note that ${A}(0, \omega)$ is the Fourier transform of the magnitude squared of $s_i(t)$.
In radar, the zero-Doppler response and the zero-delay response are used to quantify the range estimation accuracy and the velocity estimation accuracy, respectively \cite{richards14}. This is further discussed using numerical examples in Section \ref{NR}.

\subsection{Fisher Information Matrix}
For the $i$-th transmitted waveform, we define the complex envelope of the received radar waveform as given in \cite[eq. 3]{vantrees01}
\begin{align}\label{eq321}
r(t) = \tilde{b}s_i(t - \tau)e^{j\omega t} + n(t),
\end{align}
where $\tau$ and $\omega$ are the unknown delay and Doppler to be estimated, the multiplier $\tilde{b}$ represents fading induced by multiple reflecting surfaces on the target and is modelled by a zero mean and unit variance complex Gaussian random variable and $n(t) \sim \mathcal{CN}(0, N_0)$ represents the AWGN. Then, the $2 \times 2$ Fisher information matrix for $\tau$ and $\omega$ is defined as \cite{vantrees01}
\begin{align}\label{eq32}
{\bf {J}} = \begin{bmatrix}
    J_{11}  &  J_{12} \\
     J_{21} &  J_{22},
\end{bmatrix}
\end{align}
where we identify subscript 1 with $\tau$ and subscript 2 with $\omega$. Based on \cite[eq. (63) - (65)]{vantrees01}
 we can write the elements of  ${\bf {J}}$ as
\begin{align}\label{eq33}
J_{11} = C[\overline{\omega^2} - (\bar{\omega})^2],
\end{align}
\begin{align}\label{eq34}
J_{12} = J_{21} = C[\overline{\omega\tau} - \bar{\omega}\bar{\tau}],
\end{align}
\begin{align}\label{eq35}
J_{22} = C[\overline{\tau^2} - (\bar{\tau})^2],
\end{align}
where  $C = \frac{2}{N_0}\left(\frac{1}{1 + N_0}\right)$ and $\overline{\omega^2}$, $\overline{\omega\tau}$ and $\overline{\tau^2}$ are given in \cite[eq. (67), (68) and (69)]{vantrees01}, respectively. The term $\overline{\omega}$ can be found by replacing $\omega^2$ by $\omega$ in \cite[eq. (67)]{vantrees01} and similarly $\overline{\tau}$ can be found by replacing $u^2$ by $u$ in \cite[eq. (69)]{vantrees01}. Using lengthy mathematical manipulations, we solve \eqref{eq33} - \eqref{eq35} and find closed-form expressions for the Fisher information matrix elements as
\begin{align}\label{eq36}
J_{11} \approx \frac{2BC}{T} \left(1 - \left(\frac{M-1}{M}\right) \cos{\left(\omega_0T\right)}\right),
\end{align}
\begin{align}\label{eq37}
J_{12} = J_{21} \approx   -\frac{CT^2}{2}  \sum_{m=0}^{M-1}  {(2m+1)\omega_m^i},
\end{align}
\begin{align}\label{eq38}
J_{22} = \frac{CM^2T^2}{12},
\end{align}
where $B$ is the finite filter bandwidth of the radar receiver and $\omega_0 = 2\pi f_0$.  As the derivations of the $J_{11}$, $J_{12}$ and $J_{22}$ terms involve lengthy calculations, we have presented them separately in Appendices \ref{appenB}, \ref{appenC} and \ref{appenD}, respectively. Note that the pulse width $T$ in the above expressions are that of the perfectly rectangular pulse before bandwidth limiting. It is a good approximation to
the width of the bandwidth-limited pulse for large $BT$. Also, the approximation in \eqref{eq36} comes from neglecting the small terms in the resulting integrals which vanish as we set $BT \rightarrow \infty$. The term $\omega_0T$ represents the phase shift between adjacent frequency tones and is further explained in Appendix \ref{appenB}.

 Based on the Fisher information matrix we can bound the variance of delay and Doppler estimation errors which directly corresponds to the target range and velocity estimation accuracy in radar. More specifically, the variance of any unbiased estimate is bounded by the diagonal elements of ${\bf {J}}^{-1}$ \cite{vantrees01}. Thus, the CRLBs on the delay estimation error and the Doppler estimation error, which we denote by $\textrm{CRLB}_{\tau}$ and $\textrm{CRLB}_{\omega}$, respectively, can be approximated as
\begin{align}\label{eq40}
\textrm{CRLB}_{\tau} \approx C^{-1}\left(\frac{M^2T}{2M^2B\left(1 - \left(\frac{M - 1}{M}\right)\cos{(\omega_0T)}\right) - 3T^3\left(\sum_{m = 0}^{M-1}(2m+1)\omega_m^i\right)^2}\right),
\end{align}
\begin{align}\label{eq42}
\textrm{CRLB}_{\omega} \approx C^{-1}\left(\frac{\frac{2B}{T}\left(1 - \left(\frac{M - 1}{M}\right)\cos{(\omega_0T)}\right)}{\frac{M^2BT}{6}\left(1 - \left(\frac{M - 1}{M}\right)\cos{(\omega_0T)}\right) - \frac{T^4}{4}\left(\sum_{m = 0}^{M-1}(2m+1)\omega_m^i\right)^2} \right).
\end{align}
We can further simplify \eqref{eq40} and \eqref{eq42} based on \cite[eq. (94),  (95)]{vantrees01} by ignoring the effect of the $\overline{\omega\tau}$ term on the CRLB. This simplification allows us to draw more insights on the effect of different parameters on the estimation errors as follows
\begin{align}\label{eq43}
\textrm{CRLB}_{\tau} \approx C^{-1}\left(\frac{MT}{2B\left(M - (M - 1)\cos{(\omega_0T)}\right)}\right),
\end{align}
\begin{align}\label{eq44}
\textrm{CRLB}_{\omega} \approx C^{-1}\left( \frac{12}{M^2T^2} \right).
\end{align}
Note that when we set $M = 1$, the right hand sides of \eqref{eq43} and  \eqref{eq44} reduce to $C^{-1}T/2B$ and $12C^{-1}/T^2$, respectively, which are the CRLBs on the delay and Doppler estimation errors for a simple pulse \cite{richards14,skolnik}.
Based on \eqref{eq43} we note that the bound on the delay estimation error is  proportional to the simple pulse length $T$. 
In other words, the delay estimation error is inversely proportional to the effective bandwidth of the simple pulse (which is approximately $1/T$).
Based on \eqref{eq44} we note that the bound on the Doppler estimation accuracy is determined by the total pulse width. More specifically, the Doppler estimation error decreases as we increase $M$ and $T$. If we keep the energy of the signal fixed, and increase $T$ the accuracy in Doppler will increase and the accuracy in delay will decrease.
These observations are further discussed in Section \ref{NR} using numerical examples.

\subsection{Discussion}\label{disc}
As is the case with any radar waveform with a frequency sweep across the waveform, such as the LFM waveform, the Lehmer code based random stepped frequency radar waveform introduces range-Doppler coupling. Thus, when both the delay and Doppler are unknown, there is an ambiguous region in the delay-Doppler plane. 
In radar, the classical method of resolving this ambiguity is by transmitting another waveform with the opposite frequency sweep \cite{vantrees01}. In our joint radar communication system, this is automatically taken care of by the randomization of the incoming data, because, the transmitted radar waveform will differ based on the communication data, and hence, the effect of the overall range-Doppler coupling will be negligible in average. More specifically, the off-diagonal terms in the Fisher information matrix, i.e., the $J_{12}$ and $J_{21}$ terms in \eqref{eq37} change with  $\omega_m^i$, and hence on the particular frequency permutation of the waveform. They are maximized when $\omega_m^i$ values  are set in ascending order and the resulting  ambiguous region lies in the first and third quadrants of the delay-Doppler plane. Similarly, $J_{12}$ and $J_{21}$ terms are minimized when $\omega_m^i$ values are set in descending order and the resulting  ambiguous region lies in the second and fourth quadrants of the delay-Doppler plane.


The overall structure of the AF resulting from the Lehmer code based random stepped frequency radar waveform  has a narrow main-lobe and small side-lobes. Such an AF is important to achieve good radar performance specially with clutter. There is a useful averaging effect on the AF side-lobes which occurs because the communications tends to randomly select among all possible frequency combinations  which means that clutter enters via the average AF side-lobe level and this is reasonably small for our waveform.
The positioning of these side-lobes in the range-Doppler plane changes with the transmitted radar waveform. 
In situations where AF side-lobe clutter must be minimized,  a more advanced coding scheme can be used in which the incoming data is mapped to radar waveforms in such a way that the overall effect of clutter is minimized in average. This could be a topic of  interest for future.

\section{Numerical Examples}\label{NR}

\begin{figure}[t]
    \centerline{\includegraphics[width=0.6\textwidth]{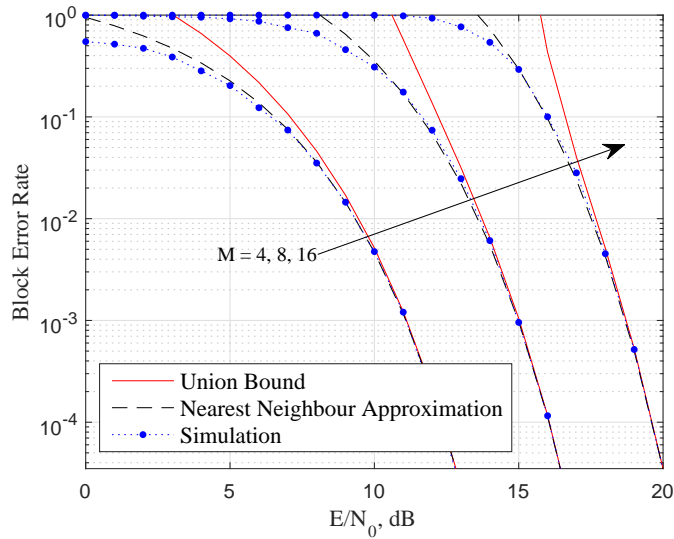}}
    \caption{The block error rate versus average received SNR for $N = 2$ and $M = 4, 8, 16$. The results are for the proposed baseband signaling model in AWGN.
    }\label{fig3}
\end{figure}

\begin{figure}[t]
    \centerline{\includegraphics[width=0.6\textwidth]{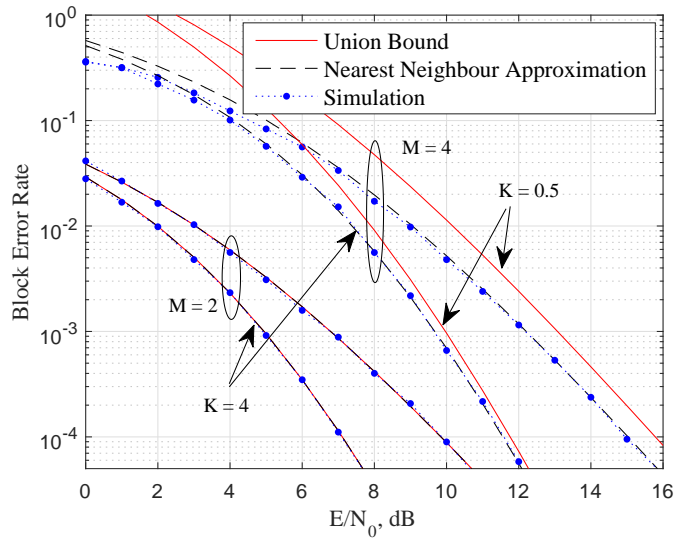}}
    \caption{The block error rate versus average received SNR for $N = 4$, $M = 2, 4$ and $K = 0.5, 4$. The results are for the proposed baseband signaling model in a Rician fading channel.
    }\label{fig4}
\end{figure}

\begin{figure}[t]
    \centerline{\includegraphics[width=0.6\textwidth]{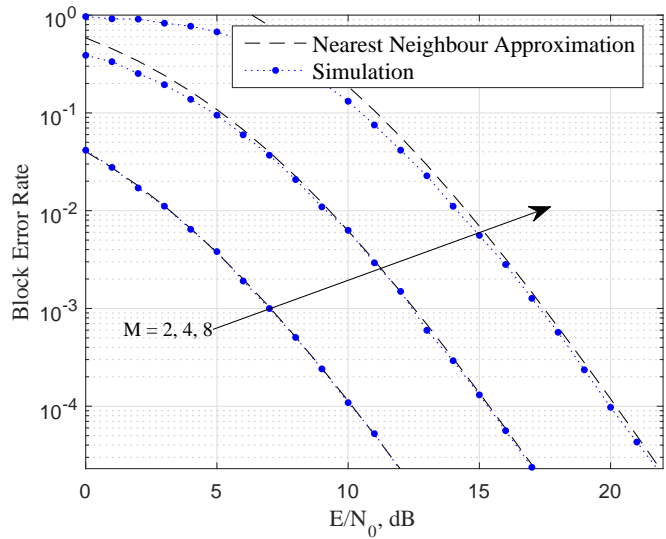}}
    \caption{The block error rate versus average received SNR for $N = 4$ and $M = 2, 4, 8$. The results are for the proposed baseband signaling model in a Rayleigh fading channel.
    }\label{fig5}
\end{figure}

\begin{figure}[t]
    \centerline{\includegraphics[width=0.6\textwidth]{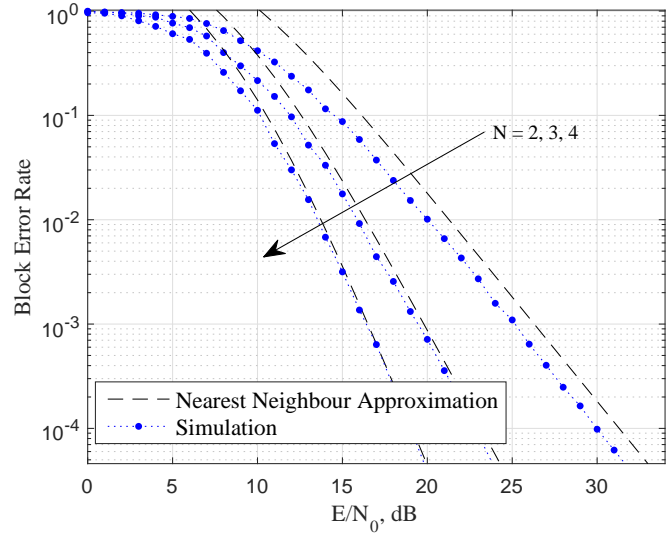}}
    \caption{The block error rate versus average received SNR for $K = 1$, $M = 8$ and $N = 2, 3, 4$. The results are for the proposed baseband signaling model in Rician fading channel.
    }\label{fig6}
\end{figure}

\begin{figure}[t]
    \centerline{\includegraphics[width=17cm,height=7cm]{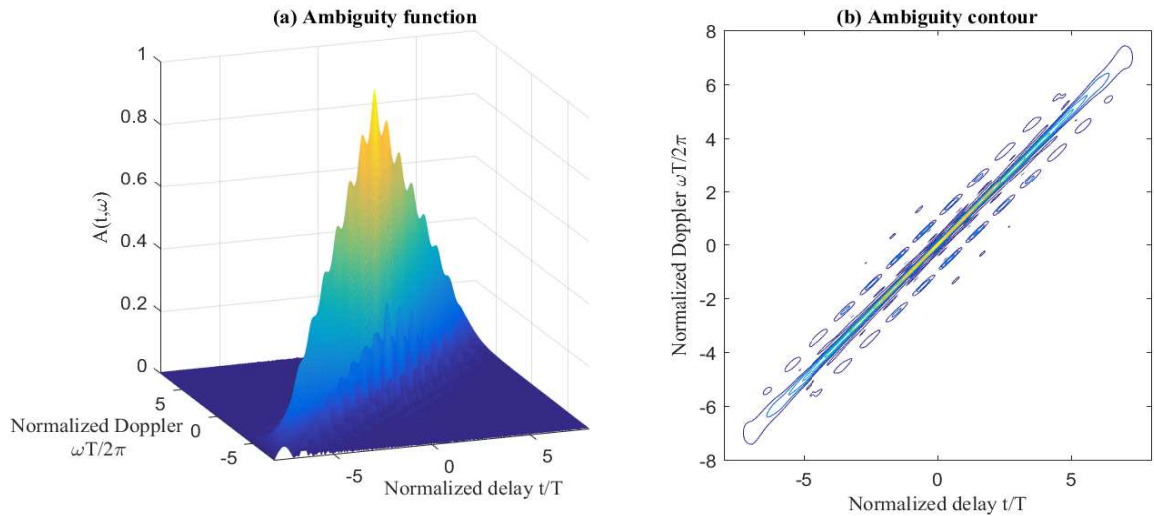}}
    \caption{Ambiguity function of a Lehmer code based random stepped frequency radar waveform for M = 8. The permutation of the frequency sequence is arranged in ascending order.
    }\label{fig7}
\end{figure}

\begin{figure}[t]
    \centerline{\includegraphics[width=17cm,height=7cm]{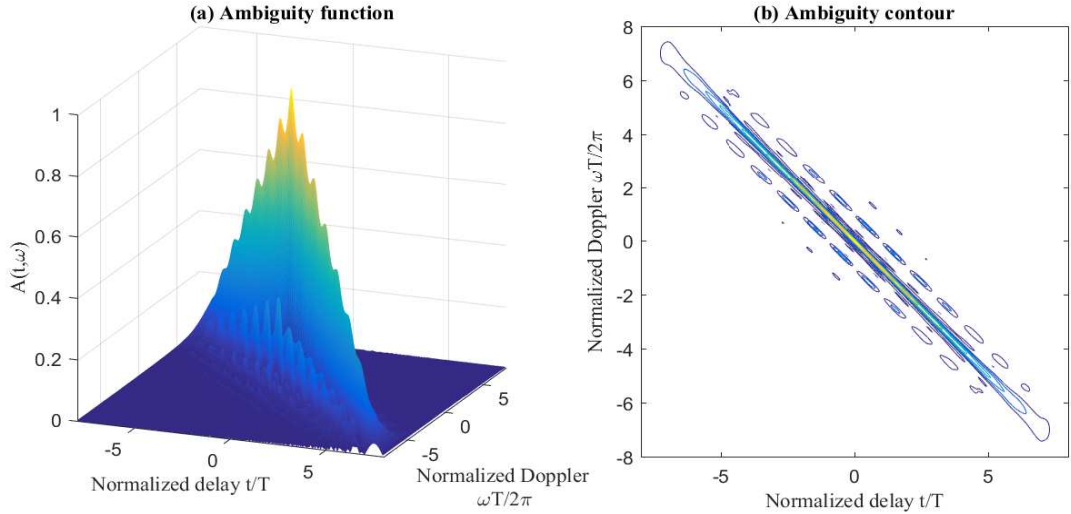}}
    \caption{Ambiguity function of a Lehmer code based random stepped frequency radar waveform for M = 8. The permutation of the frequency sequence is arranged in descending order.
    }\label{fig8}
\end{figure}

\begin{figure}[t]
    \centerline{\includegraphics[width=0.6\textwidth]{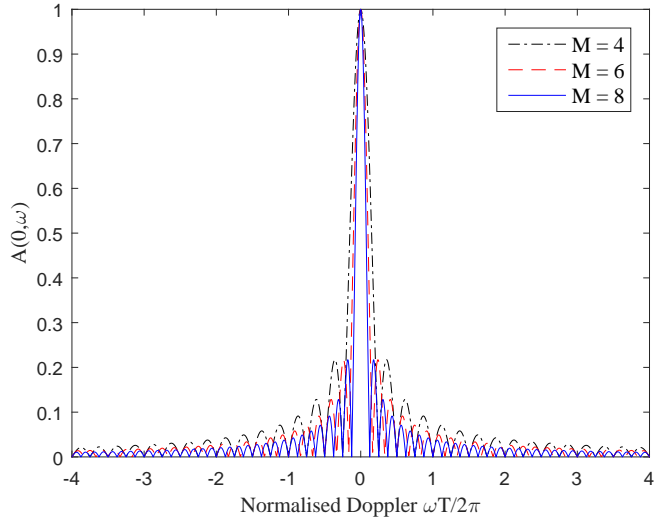}}
    \caption{The zero-delay cut of the ambiguity function for M = 4, 6, 8.
    }\label{fig9}
\end{figure}

\begin{figure}[t]
    \centerline{\includegraphics[width=0.6\textwidth]{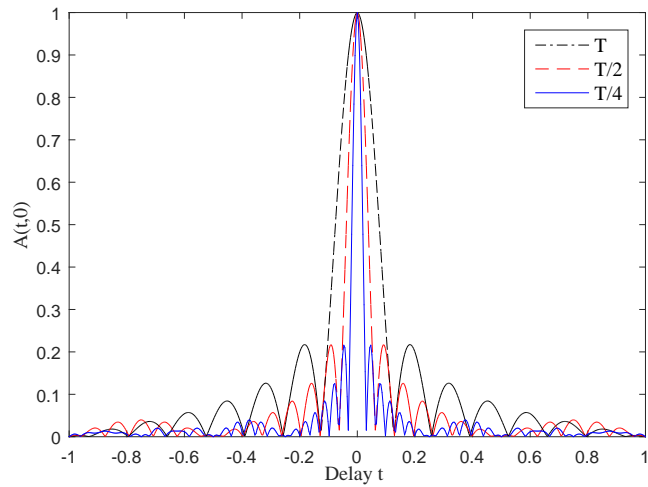}}
    \caption{The zero-Doppler cut of the ambiguity function for simple pulse width = $T, \frac{T}{2}$ and $\frac{T}{4}$.
    }\label{fig11}
\end{figure}

In this section we present numerical examples illustrating the performance of the  Lehmer code based random stepped frequency radar waveform and the proposed baseband signaling model.  Fig. \ref{fig3} - Fig. \ref{fig6} focus on the communication functionality while Fig. \ref{fig4} - Fig. \ref{fig11} focus on the radar functionality.

Fig. \ref{fig3} plots the block error rate versus the average received signal-to-noise ratio (SNR) in an AWGN channel with two diversity branches, i.e., $N = 2$, and $M = 4, 8$ and 16. The frequencies are chosen to maintain orthogonality, i.e., the frequency separation $\Delta f = 1/T$. Note that the waveform energy is assumed to be one and is kept constant as we increase $M$. The simulation results are generated using Monte-Carlo simulations while the analytical approximations are generated using the union bound in \eqref{eq19d} and the nearest neighbour approximation in \eqref{eq20d}. Both the union bound and the nearest neighbour approximation sit above the simulation curves and accurately approximate the block error rate at the high SNR regime. Since the waveform energy is kept constant the block error rate increases as we increase $M$.

Fig. \ref{fig4} plots the block error rate versus the average received SNR under the Rician fading channel model with a Rician $K$ factor 0.5 and 1. We set the number of antennas $N = 4$ and change $M = 2, 4$. Similar to Fig. \ref{fig3}, the simulation results are generated using Monte-Carlo simulations while the analytical approximations are generated using the union bound in \eqref{eq19c} and the nearest neighbour approximation in \eqref{eq20c}. As we increase the Rician $K$ factor from 0.5 to 4, the communication channel experiences a stronger LOS signal, as a result of which, the block error probability performance improves.  For $M = 2$, the  union bound and the  nearest neighbour approximation reduces to the exact block error probability, and hence coincide with the simulation results. We note that the nearest neighbour approximation accurately follows the simulated results generated using the Monte-Carlo simulations, especially in the high SNR regime. As we increase $M$ the union bound becomes quite loose. This is expected as the union bound takes into account all the pairwise error probabilities which is a significant number as we increase $M$. Therefore, in the next examples we only illustrate the nearest neighbour approximation which provides a better approximation of the block error probability than the union bound.

Fig. \ref{fig5} plots the block error rate versus the average received SNR under the Rayleigh fading channel model ($K = 0$). We set the number of antennas $N = 4$ and change $M = 2, 4, 8$ while keeping the energy of the waveform at one. The analytical approximations are generated using the the nearest neighbour approximation in \eqref{eq22}. We observe that \eqref{eq22} provides an accurate approximation of the block error probability under Rayleigh fading channels. Similar to Fig. \ref{fig3}, since the waveform energy is kept constant the block error rate increases as we increase $M$.

Fig. \ref{fig6} plots the block error rate versus the average received SNR by changing the number of antennas as $N = 2, 3$ and $4$. We consider the Rician fading channel model and set $K = 1$ and $M = 8$. The analytical results are generated using the  nearest neighbour approximation in \eqref{eq20c}. As expected, when we increase the number of receive antennas the block error rate decreases due to the receive diversity gain. We also observe that the nearest neighbour approximation  becomes more accurate with increasing $N$.

Next we focus on the radar performance of the waveform.
Fig. \ref{fig7} and Fig. \ref{fig8} plot the ambiguity function of the proposed Lehmer code based random stepped frequency radar waveform for two different permutations of $M = 8$ frequency tones. In Fig. \ref{fig7} the frequency tones are in ascending order while in Fig. \ref{fig8} they are in descending order. Subplots (a) and (b) present the three-dimensional surface plot and the contour plot, respectively. The energy of the signal is normalized so that the ambiguity function has a maximum value of one. Note that, as with any waveform the maximum of the ambiguity function occurs at $t = \omega = 0$. Around the origin, the ambiguity function rarely changes as we  change the permutation of the frequency sequence, thus the local accuracy remains the same irrespective of the transmitted waveform. The broader structure of the ambiguity function, however, can change depending on the permutation. This can be clearly observed by the contour plots given in Fig. \ref{fig7} and Fig. \ref{fig8}. Similar to the traditional LFM waveform, the ambiguity function of the Lehmer code based random stepped frequency radar waveform is skewed in the delay-Doppler plane giving rise to range-Doppler coupling. As discussed in Section \ref{disc}, this behaviour is expected as our waveform is sweeping across a bandwidth of $M \Delta f$. As shown in subplot (b) of Fig. \ref{fig7}, when the permutation of the frequency sequence is ordered in ascending order, the  ambiguous region lies in the first and third quadrants. When the permutation of the frequency sequence is ordered in descending order, the  ambiguous region lies in the second and fourth quadrants.

More insight into the ambiguity function can be obtained from the one-dimensional cuts through the delay and the Doppler axis. 
Fig. \ref{fig9} plots the zero-delay cut of the ambiguity function for $M = 4, 6$ and 8 while keeping the energy of the waveform constant at one. As we increase $M$, more frequency tones are included in the waveform which results in different side-lobe structures. Overall the waveform  achieves reasonable side-lobe performance.
Around the origin, the curvature of the main-lobe of the zero-delay cut increases as we increase $M$, thus improving the accuracy of the Doppler estimation.
This also agrees with the analytical relationship we observed in \eqref{eq44} using the approximation on the CRLB of the Doppler estimation error, which decreases as we increase the total pulse width.

Fig. \ref{fig11} plots  the zero-Doppler cut of the ambiguity function by fixing $M = 4$ and changing the pulse width as $T, T/2$ and $T/4$ while keeping the energy of the waveform constant at one. As we change the pulse width, $\Delta f$ was changed accordingly to maintain the orthogonality between frequency tones such that $T \Delta f = 1$. From the plot we note that around the origin, the curvature of the main-lobe of the zero-Doppler cut increases as we decrease the pulse width, thus improving the accuracy of the range estimation. This also agrees with the analytical relationship  we observed in \eqref{eq43} using the approximation on the CRLB on the delay estimation error, which decreases as we decrease the pulse width.

\begin{figure}[t]
    \centerline{\includegraphics[width=0.6\textwidth]{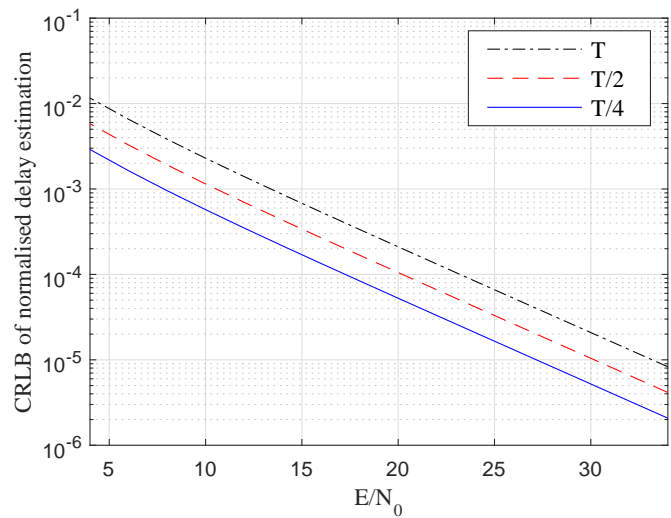}}
    \caption{Behavior of  the approximated delay  CRLB in \eqref{eq43}.
    }\label{fig12}
\end{figure}

\begin{figure}[t]
    \centerline{\includegraphics[width=0.6\textwidth]{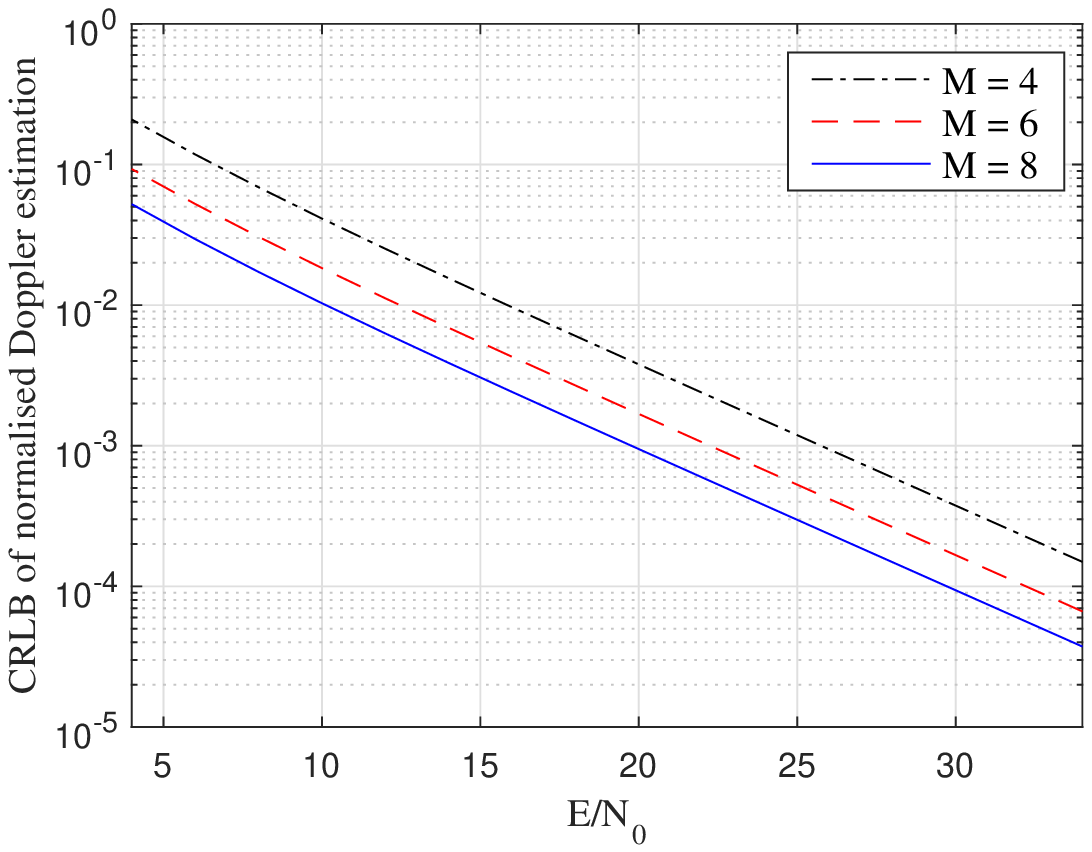}}
    \caption{Behavior of  the approximated Doppler  CRLB in  \eqref{eq44}.
    }\label{fig14}
\end{figure}

Finally, Figs. \ref{fig12} and \ref{fig14} illustrate the behaviour of the CRLBs of delay and Doppler estimation errors derived in \eqref{eq43} and \eqref{eq44}, respectively. Similar to Fig. \ref{fig11}, in Fig. \ref{fig12} we set $M = 4$ and change the pulse width as $T, T/2$ and $T/4$ while keeping the energy of the waveform constant at one. The figure illustrates the behaviour of the approximated CRLB on the delay estimation error, versus the received SNR. As we decrease the pulse width the delay CRLB decreases making a more accurate delay estimation. Similar to Fig. \ref{fig9}, in \ref{fig14} we consider $ M = 4, 6$ and 8 while keeping the energy of the waveform constant at one. As we increase $M$ the Doppler CRLB decreases making a more accurate Doppler estimation.

%

\section{Conclusion and Future Work}\label{con}
A Lehmer code based random stepped frequency radar waveform is presented as a suitable candidate for simultaneous data transmission and radar sensing. A signaling scheme is proposed in which data is embedded based on the selection of the frequency permutation in the transmitted waveform. From the data  communication perspective, the union bound and the nearest neighbour approximation on the block error probability is analysed based on the optimum ML detector. Closed-form expressions are derived for  the block error probability under the baseline case of AWGN channels and the more general case of Rician fading channels. Based on the Hungarian algorithm an efficient implementation of the communications receiver is also presented. From a radar sensing perspective, the performance of the radar waveform is fully characterized based on the ambiguity function and Fisher information matrix. CRLBs on delay and Doppler estimation errors are also derived to analyse the accuracy of range and velocity resolution. Simulation results demonstrate the accuracy of the analytical derivations and illustrate the communication and radar performance of the proposed waveform. In the following, we outline several possible directions for further research.
\begin{itemize}
  \item The communication performance in this paper focused on the block error rate. It would be desirable to formalise the extension to bit and symbol error rates which will require the selection of a subset of waveforms from the set of $M!$ waveforms. This will bring in important aspects relating to linear block codes and numbering permutations.
  \item Related to the above point, it would also be desirable to understand the tradeoffs between the achievable communication data rates and the error probability, which directly relates to the minimum distance between selected waveforms.
  \item The analysis could be extended to randomize the phase of each frequency tone, on top of the frequency permutation. This will increase the communication data rate and allow the incorporation of traditional modulation methods such as $M$-ary Phase Shift Keying (MPSK).
  \item Another desirable extension would be to relax the constraint on frequency permutations and allow the frequencies to repeat within the waveform. This will also increase the communication data rate, but the analysis of its effect on the radar performance is nontrivial.
\end{itemize}

\begin{appendices}

\section{Derivation of $\mathbf{h}^H\mathbf{h}$ in \eqref{eq13}}\label{appenA}
Using \eqref{eq12} we can express $\mathbf{h}^H\mathbf{h}$ as a summation   given by
\begin{align}\label{eqa2}
\mathbf{h}^H\mathbf{h} = \sum_{i = 1}^N \left|\sqrt{\frac{K}{K+1}}\Delta_i + \sqrt{\frac{1}{K+1}}u_i\right|^2.
\end{align}
Note that  $|\Delta_i|^2 = 1$. We can write an expression that is statistically identical to $\mathbf{h}^H\mathbf{h}$ as
\begin{align}\label{eqa3}
\mathbf{h}^H\mathbf{h} \overset{d}{=} \sum_{i = 1}^{2N} \left|\delta_i + \frac{\gamma_i }{\sqrt{K + 1}}\right|^2,
\end{align}
where $X \overset{d}{=} Y$ implies the statistical equivalence when $X$ has the same distribution as $Y$, $\gamma_i = \textrm{Re}(u_i)$ for all $i \in \{1, 2, \ldots, N\}$, $\gamma_i = \textrm{Im}(u_i)$ for all $i \in \{N + 1, N + 2, \ldots, 2N\}$, $\delta_i = \sqrt{\frac{K}{K + 1}}$ for all $i \in \{1, 2, \ldots, N\}$ and $\delta_i  = 0$ for all $i \in \{N + 1, N + 2, \ldots, 2N\}$.
Using some mathematical manipulations \eqref{eqa3} can be written as
\begin{align}\label{eqa4}
\mathbf{h}^H\mathbf{h} \overset{d}{=} \frac{1}{2(K + 1)}\sum_{i = 1}^{2N} \left|\delta_i \sqrt{2(K + 1)} + {\tilde{\gamma}_i }\right|^2,
\end{align}
where $\tilde{\gamma}_i = \sqrt{2}\gamma_i \sim \mathcal{N}(0, 1).$ Thus, we have a sum of squares of $2N$ Gaussian random variables resulting in a non-central chi-squared distribution given by
\begin{align}\label{eqa5}
\mathbf{h}^H\mathbf{h} \overset{d}{=} \frac{1}{2(K + 1)} \chi_{2N}^2(2NK).
\end{align}

\section{Working example of the Hungarian Algorithm}\label{appenAA}
\begin{figure}[t]
    \centerline{\includegraphics[width=17cm,height=3cm]{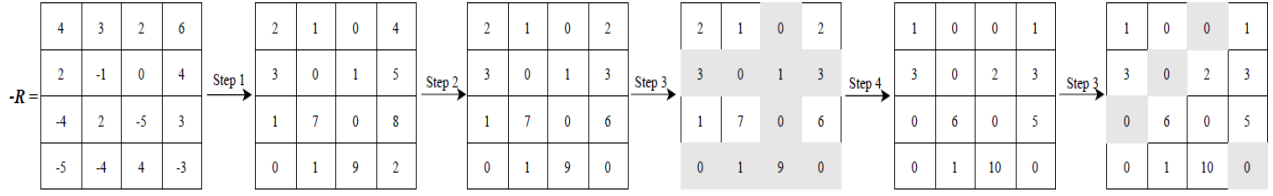}}
    \caption{An example of the Hungarian algorithm.} \label{fig:hun}
\end{figure}

In this appendix we have provide a working example of the Hungarian algorithm used in the communications receiver.
Suppose the number of receiving antennas $N = 4$. We consider the following example in which the matrix $\boldsymbol{R}$ is given by
\begin{equation}
\boldsymbol{R} = \left[\begin{array}{cccc}
    -4 & -3 & -2 & -6\\
    -2 & 1 & 0 & -4 \\
    4 & -2 & 5 & -3 \\
    5 & 4 & -4 & 3
\end{array} \right].
\end{equation}
To select $4$ elements that are in $4$ different rows and $4$ different columns and the sum of the $4$ elements is maximised, we apply the Hungarian algorithm to $-\boldsymbol{R}$. The steps of the algorithm are illustrated in Fig. \ref{fig:hun} and explained as follows.

In the first step we subtract the row minimum from each row, which results in the second matrix illustrated in Fig. \ref{fig:hun}. 
In the second step we subtract the column minimum from each column, which results in the third matrix illustrated in Fig. \ref{fig:hun}.
In the third step we determine the minimum number of lines that are required to cover all zeros in the matrix. In this example, all zeros can be covered using $3$ lines drawn across the third column, the second row and the fourth row as indicated in grey shading in the fourth matrix of Fig. \ref{fig:hun}.
Since the number of lines required $3$ is lower than the size of the matrix $4$, we continue with the fourth step. We subtract the smallest uncovered number $1$ from all uncovered elements and add it to all elements that are covered twice, which results in the fifth matrix illustrated in Fig. \ref{fig:hun}.
Then again we return to the third step. The minimum number of lines that are required to cover all zeros in the matrix is 4, which is equivalent to the size of the matrix. Therefore, an optimal assignment exists among the zeros in the matrix. The zeros cover an optimal assignment and  the resulting frequency sequence is $\{f_2, \, f_1, \, f_0, \, f_3 \}$.


\section{Derivation of the $J_{11}$ term}\label{appenB}
Let us first focus on $\overline{\omega^2}$ which is defined as
\begin{align}\label{eq1A}
\overline{\omega^2} = \int_{-\infty}^\infty \omega^2 |S_i(j\omega)|^2 \frac{d\omega}{2\pi},
\end{align}
where $S_i(j\omega)$ is the Fourier transform of $s_i(u)$ and can be derived as
\begin{align}\label{eq2A}
S_i(j\omega) = \sqrt{\frac{T}{M}}\sum_{m = 0}^{M-1} \textrm{sinc}\left((\omega_m^i - \omega)\frac{T}{2}\right)e^{-j\left(\omega mT - (\omega_m^i - \omega)\frac{T}{2}\right)}.
\end{align}
Substituting \eqref{eq2A} into \eqref{eq1A} we write
\begin{align}\label{eq3A}
\overline{\omega^2} = \frac{T}{2\pi M} \sum_{m = 0}^{M-1} \sum_{n = 0}^{M-1}e^{j(\omega_m^i - \omega_n^i)\frac{T}{2}} \int_{-\infty}^\infty\omega^2 \textrm{sinc}\left((\omega - \omega_m^i)\frac{T}{2}\right) \textrm{sinc}\left((\omega - \omega_n^i)\frac{T}{2}\right) e^{j\omega T(m-n)} d\omega.
\end{align}
Note that $\overline{\omega^2}$ does not converge. This is because we have assumed a perfect rectangular pulse, i.e, one with zero rise and zero fall time, for $s_p(u)$ in $s_i(u)$. In practice, however, pulses are not perfectly rectangular because a zero rise time or a zero fall time requires an
infinite bandwidth. Therefore, similar to \cite{skolnik} we assume that the spectrum of $s_p(u)$ has the form of a sinc function but the bandwidth is limited to a finite value of $B$ such that the integral in \eqref{eq3A} becomes
\begin{align}\label{eq5A}
\overline{\omega^2} = \frac{T}{2\pi M} \sum_{m = 0}^{M-1} \sum_{n = 0}^{M-1}e^{j(\omega_m^i - \omega_n^i)\frac{T}{2}}\int_{\omega_n^i-B\pi}^{\omega_m^i+B\pi}\omega^2 \textrm{sinc}\left((\omega - \omega_m^i)\frac{T}{2}\right) \textrm{sinc}\left((\omega - \omega_n^i)\frac{T}{2}\right) e^{j\omega T(m-n)} d\omega.
\end{align}
This is equivalent to passing a perfectly rectangular pulse through a filter of width $B$.
Substituting $v = \frac{\omega T}{2}$, $v_m = \frac{\omega_m^i T}{2}$ and $v_n = \frac{\omega_n^i T}{2}$, where we have dropped the superscript $i$ for notational simplicity, we can reexpress \eqref{eq5A} as
\begin{align}\label{eq6A}
\overline{\omega^2} = \frac{4}{\pi MT^2} \sum_{m = 0}^{M-1} \sum_{n = 0}^{M-1}e^{j(v_m - v_n)}\int_{v_n-\frac{B\pi T}{2}}^{v_m+\frac{B\pi T}{2}}v^2 \textrm{sinc}\left(v - v_m\right) \textrm{sinc}\left(v - v_n\right) e^{j2vk}dv,
\end{align}
where $k = m-n$. Next we use the  identities
\begin{align}\label{eq7A}
\frac{v^2}{(v-v_m)(v-v_n)} = 1 + \frac{(v_m + v_n)}{v - v_m} + \frac{v_n^2}{(v - v_m)(v - v_n)},
\end{align}
and
\begin{align}\label{eq8A}
\sin{(v - v_m)}\sin{(v - v_n)}e^{j2vk} = \frac{1}{2}\cos{(v_n - v_m)e^{j2vk}} -  \frac{1}{4}e^{j(2(k - 1)v + v_m + v_n)} -  \frac{1}{4}e^{j(2(k + 1)v - v_m - v_n)},
\end{align}
to reexpress \eqref{eq6A} as
\begin{align}\label{eq9A}
\overline{\omega^2} = \frac{4}{\pi MT^2} \sum_{m = 0}^{M-1} \sum_{n = 0}^{M-1}e^{j(v_n - v_m)}J_{n,m},
\end{align}
where
\begin{align}\label{eq10A}
J_{n,m} &= \int_{v_n-\frac{B\pi T}{2}}^{v_m+\frac{B\pi T}{2}}  \left(1 + \frac{(v_m + v_n)}{v - v_m} + \frac{v_n^2}{(v - v_m)(v - v_n)}\right)
\left(\frac{1}{2}\cos{(v_n - v_m)e^{j2vk}} \right.\notag\\&\left.-  \frac{1}{4}e^{j(2(k - 1)v + v_m + v_n)} -  \frac{1}{4}e^{j(2(k + 1)v - v_m - v_n)}\right) dv.
\end{align}
It is important to note that all of the integrals in \eqref{eq10A} are small and of order one when $k \neq 0, \pm 1$.  This is because for $q \neq 0$, the integrals $\int e^{jqv} dv$, $\int \frac{e^{jqv}}{v - v_m} dv$ and $\int \frac{e^{jqv}}{(v - v_m)(v - v_n)} dv$ for $m \neq n$ can all be computed using trigonometric integrals \cite[eq. (2.641.1)-(2.641.2)]{grashteyn07} giving order one integrals. Hence, the dominant terms in  $J_{n,m}$ are $k = 0$ (i.e., $n = m$) and  $k = \pm 1$ (i.e., $n = m \pm 1$) which can be written as
\begin{align}\label{eq11A}
J_{n,n} = \frac{1}{4}\int_{v_n-\frac{B\pi T}{2}}^{v_m+\frac{B\pi T}{2}}\left(1 + \frac{2v_n}{v - v_n}+\frac{v_n^2}{(v - v_n)^2}\right)\left(2 - e^{j(-2v+2v_n)} - e^{j(2v - 2v_n)}\right)dv
\end{align}
\begin{align}\label{eq12A}
J_{n,n+1} = -\frac{1}{4}e^{j(v_n + v_{n+1})}\int_{v_n-\frac{B\pi T}{2}}^{v_m+\frac{B\pi T}{2}}\left(1 + \frac{v_n + v_{n+1}}{v - v_{n+1}}+\frac{v_n^2}{(v - v_n)(v - v_{n + 1})}\right)dv + \mathcal{O}(1)
\end{align}
\begin{align}\label{eq13A}
J_{n,n-1} = -\frac{1}{4}e^{j(v_n + v_{n-1})}\int_{v_n-\frac{B\pi T}{2}}^{v_m+\frac{B\pi T}{2}}\left(1 + \frac{v_n + v_{n-1}}{v - v_{n-1}}+\frac{v_n^2}{(v - v_n)(v - v_{n - 1})}\right)dv + \mathcal{O}(1)
\end{align}
For large $B$, we can ignore the terms involving inverse powers of $v - v_n$, $v_n - v_{n-1}$ and $v - v_{n+1}$ as these are also order one. Hence, we approximate $J_{n,n} \approx \frac{B\pi T}{2}$,  $J_{n,n+1} \approx -\frac{B\pi T}{4}e^{j(v_n + v_{n+1})}$ and $J_{n,n-1} \approx -\frac{B\pi T}{4}e^{j(v_n + v_{n-1})}$ which results in
\begin{align}\label{eq14A}
\overline{\omega^2} &\approx \frac{2B}{MT} \left({M} - \sum_{n = 0}^{M - 2}\frac{1}{2}e^{j(v_n -v_{n+1})}e^{j(v_n +v_{n+1})} - \sum_{n = 1}^{M - 1}\frac{1}{2}e^{j(v_n -v_{n-1})}e^{-j(v_n + v_{n-1})} \right).
\end{align}
After some straightforward mathematical manipulations \eqref{eq14A} can be reexpressed as
\begin{align}\label{eq15A}
\overline{\omega^2} \approx \frac{2B}{MT} \left(M - \sum_{n = 0}^{M-2} \cos{\left(\omega_n^iT\right)}\right).
\end{align}
Note that the argument inside the cosine function in \eqref{eq15A}, i.e., $\omega_n^iT$ simply corresponds to the phase change of the waveform at $T, 2T, \ldots, (M-1)T$. Since we consider orthogonal frequencies such that the separation between two frequency tones $\Delta f = n/T$, where $n$ is an integer, we can write $\omega_n^i = 2\pi (f_0 + q_n^i\Delta f)$ where $q_n^i$ is an integer and further reduce \eqref{eq15A} to derive
\begin{align}\label{eq16A}
\overline{\omega^2} \approx \frac{2B}{T} \left(1 - \left(\frac{M-1}{M}\right) \cos{\left(\omega_0T\right)}\right).
\end{align}
From \eqref{eq16A} we learn that, if we set $\omega_0T  = 0$, the cosine function takes its maximum value of one and \eqref{eq16A} reduces to
\begin{align}\label{eq17A}
\overline{\omega^2} \approx \frac{2B}{TM}.
\end{align}
Similarly, if we set $\omega_0T  = \pi$, the cosine function takes its minimum value of negative one and \eqref{eq16A} reduces to
\begin{align}\label{eq18A}
\overline{\omega^2} \approx \frac{2B}{T} \left(\frac{2M - 1}{M}\right).
\end{align}

Next, we focus on $\overline{\omega}$ which is defined as
\begin{align}\label{eq19A}
\overline{\omega} = \int_{-\infty}^\infty \omega |S_i(j\omega)|^2 \frac{d\omega}{2\pi}.
\end{align}
Following similar steps to $\overline{\omega^2}$ we write \eqref{eq19A} as
\begin{align}\label{eq20A}
\overline{\omega} = \frac{2}{\pi MT} \sum_{m = 0}^{M-1} \sum_{n = 0}^{M-1}e^{j(v_m - v_n)}\int_{v_n-\frac{B\pi T}{2}}^{v_m+\frac{B\pi T}{2}}v \textrm{sinc}\left(v - v_m\right) \textrm{sinc}\left(v - v_n\right) e^{j2vk}dv,
\end{align}
which can be reexpressed as
\begin{align}\label{eq21A}
\overline{\omega} = \frac{2}{\pi MT} \sum_{m = 0}^{M-1} \sum_{n = 0}^{M-1}e^{j(v_n - v_m)}J_{n,m},
\end{align}
where
\begin{align}\label{eq22A}
J_{n,m} &= \int_{v_n-\frac{B\pi T}{2}}^{v_m+\frac{B\pi T}{2}}  \left(\frac{v_m }{(v - v_m)(v_m - v_n)} + \frac{v_n}{(v - v_n)(v_n - v_m)}\right)
\left(\frac{1}{2}\cos{(v_n - v_m)e^{j2vk}} \right.\notag\\&\left.-  \frac{1}{4}e^{j(2(k - 1)v + v_m + v_n)} -  \frac{1}{4}e^{j(2(k + 1)v - v_m - v_n)}\right) dv.
\end{align}
Noting that all the integrals in \eqref{eq22A} are order one, for large $B$ we can approximate
\begin{align}\label{eq23A}
\overline{\omega} \approx 0.
\end{align}
Based on \eqref{eq16A} and \eqref{eq23A} we can approximate the $J_{11}$ term as given in \eqref{eq36}.

\section{Derivation of the $J_{12}$ term}\label{appenC}
Based on \eqref{eq23A} we note that $\bar{\omega} \approx 0$. Hence, we first approximate the $J_{12}$ term in \eqref{eq34} by
\begin{align}\label{eqC1}
J_{12} \approx C\overline{\omega\tau}.
\end{align}
Next, we focus on $\overline{\omega\tau}$ which is defined in \cite[eq. (68)]{vantrees01} as
\begin{align}\label{eqC2}
\overline{\omega\tau} = \textrm{Im}\left(\int_{-\infty}^\infty u s_i(u) \frac{\partial s_i^*(u)}{\partial u} du\right),
\end{align}
where $\textrm{Im}(.)$ denotes the imaginary part of the argument, $s_i(u)$ is given in \eqref{eq14} and $s_i^*(u)$ denotes the complex conjugate of $s_i(u)$. We reexpress $s_i(u)$ as
\begin{align}\label{eqC3}
s_i(u) = \sqrt{\frac{1}{MT}}\sum_{m = 0}^{M-1} s_p(u - mT)e^{2\pi j f_m^i (u - mT)},
\end{align}
Next, we approximate $s_p(u)$ by an increasing sequence of smooth
functions, $s_p^{(n)}(u)$, vanishing and  with vanishing derivatives at the endpoints $0, T$.  Specifically, we require
\begin{equation}
  \label{eq:3}
  s_p^{(n)}(u)s_p^{(n)}(u-mT)=0,
\end{equation}
and
\begin{equation}
  \label{eq:3a}
  s_p^{(n)}(u){s'}_{p}^{(n)}(u-mT)=0  for m\ne 0,\ \forall u.
\end{equation}
Let us denote the integral within the brackets in \eqref{eqC2} by $\Phi$ 
and approximate $\Phi$ by $\Phi^{(n)}$ where we  are replacing $s_p(u)$ by $s_p^{(n)}(u)$ in the formula for $s_i(u)$ and write
\begin{align}
  \label{eq:6}
  \Phi^{(n)} =  \int_{-\infty}^{\infty} &u  \Bigl(\sum_{m=0}^{M-1} s_p^{(n)}(u-mT)e^{2\pi j f_m^iu}\Bigr) \notag\\&\times \Bigl(\sum_{m'=0}^{M-1} {s'}_p^{(n)}(u-m'T)e^{-2\pi j f_{m'}^iu}
-\sum_{m'=0}^{M-1} 2\pi j f_{m'}^i s_p^{(n)}(u-m'T)e^{-2\pi j f_{m'}^iu}
  \Bigr) \, du.
\end{align}
In view of $s_p^{(n)}(u)s_p^{(n)}(u-mT)=0$, the expression in \eqref{eq:6}  becomes
\begin{equation}
  \label{eq:8}
  \begin{aligned}
\Phi^{(n)} =\sum_{m=0}^{M-1}\Bigl(\int_{-\infty}^{\infty} u    s_p^{(n)}(u-mT) \bigl({s'}_p^{(n)}(u-mT)\,du -2\pi j f_{m}^i\int_{-\infty}^{\infty} u   s_p^{(n)}(u-mT)^2\,du\Bigr).
\end{aligned}
\end{equation}
Focusing just on the imaginary part of $\Phi^{(n)}$ we can write an approximation for $ \overline{\omega\tau}$ as
\begin{align}\label{eq:9}
   \overline{\omega\tau} &\approx  -2\pi \sum_{m=0}^{M-1} f_{m}^i \int_{-\infty}^{\infty} u   s_p^{(n)}(u-mT)^2\,du \notag\\ 
   &=-\pi\sum_{m=0}^{M-1}  f_{m}^i {(2m+1)T^2}.
\end{align}
Substituting \eqref{eq:9} into \eqref{eqC1}  we can derive an approximation for  $J_{12}$ as
\begin{equation}
  \label{eq:12}
  J_{12} \approx   -\pi C T^2  \sum_{m=0}^{M-1} (2m+1)f_m^i,
\end{equation}
which can be simplified to produce \eqref{eq37}.

\section{Derivation of the $J_{22}$ term}\label{appenD}
Let us first focus on $\bar{\tau}$ which is defined as
\begin{align}\label{eq1D}
\bar{\tau} = \int_{-\infty}^\infty u |s_i(u)|^2 du.
\end{align}
Substituting \eqref{eq14} into \eqref{eq1D} we can write
\begin{align}\label{eq2D}
\bar{\tau} = \sum_{m = 0}^{M-1}\int_{mT}^{(M+1)} \frac{u}{MT} du.
\end{align}
After some straight-forward linear algebra the above summation can be reduced to
\begin{align}\label{eq3D}
\bar{\tau} = \frac{MT}{2}.
\end{align}
Similarly, we proceed to solve  $\overline{\tau^2}$ which is defined as
\begin{align}\label{eq4D}
\overline{\tau^2} = \int_{\infty}^\infty u^2 |s_i(u)|^2 du.
\end{align}
Substituting \eqref{eq14} into \eqref{eq4D} we can write
\begin{align}\label{eq5D}
\overline{\tau^2} = \sum_{m = 0}^{M-1}\int_{mT}^{(M+1)} \frac{u^2}{MT} du,
\end{align}
which can be solved using some straight-forward linear algebra to derive
\begin{align}\label{eq6D}
\overline{\tau^2} = \frac{M^2T^2}{3}.
\end{align}
Substituting \eqref{eq3D} and \eqref{eq6D} into \eqref{eq35} the final expression of $J_{22}$ term can be derived as given in \eqref{eq38}.

\end{appendices}

\bibliographystyle{IEEEtran}
\bibliography{mybib}

\end{document}